\documentclass[pre,twocolumn, showpacs]{revtex4}
\usepackage{graphicx}
\usepackage{dcolumn}

\begin{document}
\title{Constrained dynamics of a polymer ring enclosing a constant area} 

\author{Arti Dua and Thomas A. Vilgis} 

\address{Max-Planck-Institute for
  Polymer Research, Ackermannweg 10, D-55122 Mainz, Germany}

\date{\today}

\begin{abstract}
 The dynamics of a polymer ring enclosing a constant {\sl
    algebraic} area is studied. The constraint of a constant area is found to
  couple the dynamics of the two Cartesian components of the position vector
  of the polymer ring through the Lagrange multiplier function which is time
  dependent. The time dependence of the Lagrange multiplier is evaluated in a
  closed form both at short and long times. At long times, the time dependence
  is weak, and is mainly governed by the inverse of the first mode of the
  area. The presence of the constraint changes the nature of the relaxation of
  the internal modes. The time correlation of the position vectors of the ring
  is found to be dominated by the first Rouse mode which does not relax even
  at very long times. The mean square displacement of the radius vector is
  found to be diffusive, which is associated with the rotational diffusion of
  the ring.
\end{abstract}


\maketitle

\section{Introduction}

The equilibrium statistics of a planar Brownian motion enclosing a constant
area was first studied by L{\'e}vy in 1940 with a view to obtain the
probability distribution function for the {\sl{algebraic}} area \cite{levy,levy1}. A simple analogy to topologically
constrained (entangled) polymers, which can approximately be represented by a
planar random walk constrained to enclose a constant area, was later pointed
out by Brereton and Butler \cite{brereton}. To describe the configurational
and mechanical properties of these polymers, they re-addressed this problem
using discrete Gaussian chains. An exact expression for the probability
distribution function was subsequently obtained using continuous models for
Gaussian chains \cite{khandeker,duplantier}.

Most of the studies on constrained polymer rings have been restricted
to static equilibrium behavior. However, the dynamics of a polymer ring
constrained to enclose a constant area can serve as a useful model in several
different areas of statistical physics. The Rouse dynamics of a
highly dense ensemble of long polymer chains, in general, is a simple
and well known example of the constrained dynamics due to the presence of
entanglements. The highly non-local nature of these constraints  
renders the analytical solution of the problem difficult. Therefore,
the dynamics of polymer melts is mainly studied by phenomenological tube models
\cite{doi} and their extensions \cite{macleish}. For a polymer ring,
on the other hand, analogies with topologically constrained loops can
be realized at a more formal level \cite{brereton}. The mathematical formulation of the constant area constraint is strongly related, for instance, to
the Gaussian invariance of the entangled loops \cite{doi}, and topological
problems concerning the writhe of a polymer ring
\cite{brereton1}. Some connections with rings enclosing topological
obstacles can also be discussed within a similar framework \cite{nechaev}.

As another instance, the polymer ring  can be
regarded as a very simple model to describe the dynamics of a two-dimensional
vesicle bounded by a one-dimensional lipid membrane, the perimeter of which
undergoes randomly kicked motion due to the stochastic thermal forces from the
solvent; the osmotic pressure difference between the two sides of the membrane
is such that it keeps the area constant at all times. The constraint of 
constant area can be enforced by introducing a time dependent Lagrange
multiplier in the stochastic equations for a polymer ring. The Lagrange
multiplier, therefore, plays the role of the time dependent differential
pressure, the dynamics of which ensures that the area is constant at all
times.

The technical difficulties involved in treating the constrained dynamics of a
polymer ring arise from the global nature of the area constraint. Additionally, since the constraint of a constant area has to be satisfied at each time $t$,
the dynamical behavior of the two Cartesian components of a spatial
vector ${\bf r}(s,t) = (r_{x}(s,t), r_{y}(s,t))$ at any point $s$
along the curve are no longer independent --- the rigid constraint of a
constant area strongly couples the two coordinates through the
Lagrange multiplier. The time dependence of the latter makes the analytical solution of the stochastic equations difficult.

In a more general context, the present problem belongs to the
comparatively less understood field of stochastic dynamics with rigid
constraints. Even in a more specific case of the dynamics of polymers
with rigid constraints, the literature is mainly restricted to the
Kirkwood theory for hydrodynamics and the dynamics of semiflexible polymers \cite{doi}. One reason for the
limited literature on such systems is the technical difficulty involved in
dealing with dynamical problems with rigid constraints. The nature of the
difficulty become apparent in the field-theoretic formulation of the
dynamical problem, where the presence of any constraint changes the number of
dynamical variables and adds the Lagrange multiplier to the set of variables.
Any functional formulation of the dynamics of the constrained system is,
therefore, altered by an additional Jacobian, which is
difficult to determine. In some cases the introduction of
non-commuting Grassmann variables can facilitate a solution \cite{zinn}. For the present
problem, however, we show that the nature of the area constraint is
such that a simple Langevin description suffices and a field-theoretic formulation is not neccesary.

Within the Langevin description, the area
constraint is accounted for by the Lagrange multiplier function which is time
dependent. The
time dependence of the Lagrange multiplier can be calculated in a closed form
 (which is quite unusual for the
problems of this nature) for an initial perturbation of the area modes and allows us to solve the coupled stochastic
equations for the polymer ring. It is shown that the dynamics of a polymer
ring enclosing  constant area is mainly governed by the lowest Rouse mode.
The constraint of a constant area is fixed by the first Rouse mode (associated
with rotation \cite{doi}), which is found to be diffusive at long times. The
present problem is a simple and well chosen example to describe several interesting aspects of stochastic dynamics with rigid constraints.

\section{Model}

Let us first introduce the model we are going to study and review some
of the static results. The algebraic area enclosed by a planar random walk in
a continuous representation is defined by
\begin{equation}
\label{area1}
{A}\{{\bf r}(s,t)\} =   \frac{1}{2}\int_{0}^{N}ds~ \left({\bf r}(s,t)
\times \frac{\partial{\bf r}(s,t)}{\partial s}\right) \cdot {\bf k}, 
\end{equation}  
where ${\bf k}$ is the unit vector perpendicular to the $(x,y)$ plane; ${\bf r}(s,t)$ is the position vector starting from the centre
of mass of the chain (placed at the origin) to the segment $s$ along
a chain of contour length $N$, which for a ring satisfies ${\bf
  r}(N,t) = {\bf r}(0,t)$. In statics, the constraint of a constant
area is imposed by including  a delta function $\delta({
  A}\{{\bf r}\} - {A})$ in the partition function such that the
probability distribution function for $A$ is given by
\begin{eqnarray}
\label{prob1}
P(A,N) &=& {\cal N}\int_{{\bf r}_{0}, 0}^{{\bf r}_{0}, N}{\cal D}[{\bf
  r}(s)]~ \delta({A}\{{\bf r}\} - {A})\nonumber\\
& & ~~~~~~~ \exp\left[
  -\frac{1}{b^2}\int_{0}^{N} ds \left(\frac{\partial {\bf r}(s)}{\partial s}\right)^2 \right], 
\end{eqnarray}   
where ${\cal N}$ is the normalisation constant and $b$ is the segment length; the term in the exponential is the Wiener measure which accounts for the entropic elasticity of the polymer. The area constraint restricts the conformations of
a polymer ring in such a way that closed random walks of only a given constant
area $A$ contribute to the partition function. Since the sign of the area (as
defined by Eq. (\ref{area1})) depends on the orientation of the contour, the
constraint restricts the magnitude of the area but has no control over the
shape of a ring. The latter implies that a random walk with a large number of
wiggles along its contour length will undergo many cancellations to keep the
average area constant\cite{brereton}.

The delta function in Eq. (\ref{prob1}) can be Fourier transformed, i.e.
$\delta({ A}\{{\bf r}\} - { A}) = \int dg ~\exp(ig({ A}\{{\bf r}\} - {A}))$,
such that $g$ is a variable conjugate to $A$. The probability distribution
$P(A,N)$ can then be determined for a discrete closed random walk using normal
mode analysis \cite{brereton} or its continuous representation using
functional integration \cite{khandeker,duplantier}; it is given by $P(A, N) =
(2Nb^2{\cosh}^2(2 \pi A/Nb^2))^{-1}$, which has the asymptotic
behavior of $\exp(-4 \pi A/Nb^2)$ for $A/Nb^2 \gg 1$.

The constrained dynamics of a constant area can be treated by the method of
Lagrange multipliers \cite{doi}; the idea is to add the constraining force
explicitly in the Langevin equation. The constraint of a constant area when
introduced using a Lagrange multiplier function $\lambda(t)$ produces a
coupling between $x$ and $y$ coordinates. The coupled Langevin equations are given
by
\begin{eqnarray}
\label{langevin1}
\zeta\frac{\partial {r}_{x}(s,t)}{\partial t} &=&
\frac{2k_{B}T}{b^2}\frac{\partial^2 {r}_{x}(s,t)}{{\partial s}^2} +
k_{B}T \lambda(t)\frac{\partial {r}_{y}(s,t)}{\partial s}\nonumber\\ 
& & + f_{x}(s,t) \\
\label{langevin2}
\zeta\frac{\partial {r}_{y}(s,t)}{\partial t} &=&
\frac{2k_{B}T}{b^2}\frac{\partial^2 {r}_{y}(s,t)}{{\partial s}^2} -
k_{B}T \lambda(t)\frac{\partial {r}_{x}(s,t)}{\partial s}\nonumber\\
& & + f_{y}(s,t), 
\end{eqnarray} 
where the first term in the above equations represents dissipation due to the
viscous forces of the solvent with friction coefficient $\zeta$; $f$ is the
fluctuating force whose time average is zero. The dissipation and the
fluctuations are related to each other by the simplest form of the fluctuation
dissipation theorem which dictates
$\left<f_{i}(s,t)f_{j}(s^{\prime},t^{\prime})\right> = 2\zeta
k_{B}T\delta_{ij}\delta(s-s^{\prime})\delta(t-t^{\prime})$. The second term
represents the elastic force due to the chain connectivity. The coupling
shows the most trivial effect of the constraint: the dynamics of
the two components $r_{x}(s,t)$ and $r_{y}(s,t)$ is no longer independent, as
is the case in the absence of constraint.

In terms of the normal modes, i.e. ${\bf r}_{p}(t) =
\frac{1}{N}\int_{0}^{N} ds~{\bf r}(s,t) e^{2ip\pi s/N}$, 
Eqs.(\ref{langevin1}) and (\ref{langevin2}) can be rewritten as
\begin{eqnarray}
\label{langevin3}
\zeta_{p}\frac{d {r}_{px}(t)}{dt} &=&
\frac{-8p^2\pi^2 k_{B}T}{Nb^2}{r}_{px}(t) - 2ip\pi k_{B}T
\lambda(t){r}_{py}(t)\nonumber\\
& & + f_{px}(t)\\
\label{langevin4}
\zeta_{p}\frac{d {r}_{py}(t)}{dt} &=&
\frac{-8p^2\pi^2 k_{B}T}{Nb^2}{r}_{py}(t) + 2ip\pi k_{B}T
\lambda(t){r}_{px}(t)\nonumber\\
& & + f_{py}(t),
\end{eqnarray} 
where $\zeta_{0} = N\zeta$ and $\zeta_{p\neq 0}= 2N\zeta$\cite{doi}. $A(t) = \sum_{p = -\infty}^{\infty}(-ip\pi) A_{p}(t)$, where
\begin{equation}
\label{area2}
A_{p}(t) = r_{px}(t)r_{-py}(t) -  r_{py}(t)r_{-px}(t).
\end{equation}

A more compact vectorial representation of the set of coupled Langevin
equations can be presented by using a matrix notation,
\begin{equation}
\zeta_{p} \frac{d}{dt}{\bf r}_{p}(t) = - k_{B}T{\bf M}_{p}(t)\cdot{\bf r}_{p}(t) +{\bf f}_{p}(t)
\end{equation}
where the matrix ${\bf M}_{p}(t)$ is defined
\begin{equation}
{\bf M}_{p}(t) = \pmatrix{
8p^2\pi^2 /(Nb^2) & 2ip\pi  \lambda(t) \cr
-2ip\pi  \lambda(t) & 8p^2\pi^2 /(Nb^2) \cr
},
\end{equation}
whose eigenvalues $\ell_{e}$ are given by
\begin{equation}
\ell_{e} = 8p^2\pi^2 /(Nb^2) \pm 2 p\pi  \lambda(t),
\end{equation}
which correspond to the relaxation of the individual
Rouse modes of the chain the details of which will be discussed later.  

Since the area constraint is local in time $t$ but global in the contour
length $s$, the Lagrange multiplier $\lambda$ depends only on time. The time
dependence of Lagrange multiplier makes the exact solution of the coupled
Langevin equations difficult. Any exact or approximate solution requires a
complete knowledge of the time dependence of $\lambda(t)$. Since the dynamics
of $\lambda(t)$ has to be such that the area $A(t)$ has to be constant at all
times, we begin by writing down the equation of motion for the area
modes itself, the details of which are presented below.

\section{Area Conservation}

\subsection{Dynamics of mode-dependent area}

To get a dynamical equation for the area of a polymer ring, we add and
subtract the imaginary part of Eq. (\ref{langevin4}) from Eq.
(\ref{langevin3}) in such a way that we get dynamical equations for $(r_{px} +
ir_{py})$ and $(r_{-px} - ir_{-py})$.

A series of simple steps will provide a stochastic (Langevin type) equation
for area: multiply the dynamical equation for $(r_{px} + ir_{py})$ by
$(r_{-px} - ir_{-py})$ (that is its complex conjugate) and $(r_{-px} -
ir_{-py})$ by $(r_{px} + ir_{py})$; after having added the two equations,
separate the real and the imaginary parts of the resulting equation, which for
$p\neq 0$ modes are given by the following expressions respectively:
\begin{eqnarray} 
\label{langevin7}
\zeta_{p}\frac{d {r_{p}}^2(t)}{dt} &=&
-\left(2 k p^2 + 4p\pi k_{B}T 
\lambda(t)\right){r_{p}}^2(t)\\
\label{langevin8}
\zeta_{p}\frac{d A_{p}(t)}{dt} &=&
-\left( 2 k p^2 + 4p\pi k_{B}T \lambda(t)\right)A_{p}(t), 
\end{eqnarray}  
where $k = 8\pi^2k_{B}T/Nb^2$, $A_{p}(t)$ is given by Eq. (\ref{area2}) and ${r_{p}}^2(t) =
r_{px}(t)r_{-px}(t) + r_{py}(t)r_{-py}(t)$. It is not surprising that the
above two equations have a similar structure since the first one represents
the radius and the second one the area of the $p$th mode of the ring. To get
the time dependence of $\lambda(t)$, we are interested in the time development
of the average of $A_{p}(t)$, which in thermal equilibrium is time
independent. Therefore, in writing the dynamical equation for $A_p(t)$, we
have implicitly assumed an average over a non-equilibrium ensemble in which
$\left<A_p(t)\right>_{neq}$ is explicitly time dependent. This gives
us the dynamical response of the Lagrange multiplier to a perturbation of the
area modes away from their thermal equilibrium values \cite{dieter}.

The area conservation demands $dA(t)/dt = \sum_{p}(-ip\pi)dA_{p}(t)/dt =
0$. This constraint when used in Eq. (\ref{langevin8}) produces the
following expression for the Lagrange multiplier:
\begin{equation}
\label{lm}
\lambda(t) = -\frac{4\pi}{Nb^2}\frac{\sum_{p}p^3A_{p}(t)}{\sum_{p}p^2A_{p}(t)},
\end{equation}
where $Nb^2/4\pi$ is the area of the first mode; it corresponds to a random
walk which on an average forms a circle. For a random walk to
form a circle in an average sense, the circumference $2\pi r$ should
be equal to $N^{1/2}b$, where $r$ is the radius of the circle; this condition
determines $r = N^{1/2}b/2\pi$. Thus, the area of the circle on
an average turns out to be $Nb^2/4\pi$. The second mode is a 
symmetric figure of eight and so on. 

The time dependence of the Lagrange multiplier can be determined in a
self-consistent way. To begin with we consider $\lambda(t)$ to be
independent of time and given by ${-4\pi}/{Nb^2}$. This
approximation amounts to solving Eq. (\ref{langevin8}),
which is a simple first order differential equation in time, and can
be easily solved to produce $A_{p}(t) = A_{p}(0)\exp(-t/\tau_{p})$,
where $\tau_{p}$ is the relaxation time of the $p$th mode given by $\tau_{p} =
N^2b^2\zeta/[8\pi^2k_{B}Tp(p-1)]$. The latter implies that all modes except 
the first one, which has a constant area $A_{1}(0)$, decay with the
relaxation time of $\tau_{p}$. The same is true for the radius of the
$p$th mode.

We can now substitute the expression for $A_{p}(t)$ into Eq. (\ref{lm}) to see
how $\lambda(t)$ evolves with time. In the limiting case of $\lambda(t
\rightarrow \infty) = -4\pi/Nb^2$, the dominance of the first mode is evident.
The latter is true for all times greater than the relaxation time $\tau =
N^2b^2\zeta/[8\pi^2k_{B}T] $. It is clear from Eq. (\ref{lm}) that for all
times $t \ll \tau$, the dynamics is determined by the summation over large
number of modes; for times $t \gg \tau$, on the other hand, the dynamics is
dominated by the first mode. Thus, the dynamics of $\lambda(t)$ is reflected
in its dependence on $p$ modes, which decays in time in such a way that for $t
\gg \tau $ it is only determined by $p=1$.

To get an idea of the time dependence for times much less than the relaxation
time $\tau$, one can replace the summation by an integration over $p$ modes in
Eq. (\ref{langevin8}). In the limit of $t \ll \tau$, the time dependence of
the Lagrange multiplier is approximately given by $\lambda(t) \approx -
4\pi/Nb^2 (\tau/t)^{1/2}$. It is to be noted that in carrying out the
integration over the modes, we have neglected the $p$ dependence of
$A_{p}(0)$. The mode dependence of $A_{p}(0)$, however, does not alter the way
$\lambda(t)$ scales with $t$. At short times, the nature of the decay of the
$p$th mode is reflected in the time dependence of $\lambda(t)$. It is evident
from the expression for $A_{p}(t)$ that for $t \gg \tau/p^2$, the $p$th mode
has already relaxed and does not contribute to the summation in Eq.
(\ref{lm}). This implies that the number of modes that contribute to the
summation in Eq.(\ref{lm}) at any time $t$ are approximately given by
$p^{*}(t) \approx (\tau/t)^{1/2}$. The algebraic time dependence  of
$\lambda(t)$ can easily be understood from Eq. (\ref{langevin8}). If it is
assumed that in this time regime some modes have already relaxed,
then $dA_{p^{*}}/dt = 0$ for these modes. Then the
solution for the Langrange multiplier $\lambda(t)$ is approximately given by
\begin{equation}
\lambda(t) \simeq - \frac{4 \pi}{Nb^2} p^{*}.
\end{equation}
Since the modes follow the Rouse relaxation, the inverse square root
time dependence of $\lambda(t)$ can be understood.

The time dependent expression for $\lambda(t)$ when substituted into Eq.
(\ref{langevin8}) yields $A_{p}(t) \approx A_{p}(0) {\rm e}^{-p^2t/\tau} {\rm
  e}^{p(t/\tau)^{1/2}}$, which at short times is a stretched exponential. The
latter implies that for time $t \ll \tau$, the constraint of constant area
affects the dynamics of $A_{p}(t)$ in such a way that the lower modes grow at
the expense of the higher ones leaving the total area constant; for time $t
\gg \tau$ only the first mode, which attains a constant value, survives. In
contrast, the dynamics of $\lambda(t)$ is such that it decays in time as
$(\tau/t)^{1/2}$ and at $t \gg \tau$ attains a constant value of $-4\pi/Nb^2$.

This gives us an idea about the dynamics of $\lambda(t)$, which for all times
$t \gg \tau$, is weakly dependent on time and is mainly dominated by the first
mode. In what follows, we will replace $\lambda(t)$ by a time independent
quantity $\lambda_{0}$ to discuss the statics of a closed random walk.

\subsection{Statics}

To recover some of the equilibrium results of the earlier studies from
dynamics, we begin by introducing temporal Fourier transform into Eqs.
(\ref{langevin3}) and (\ref{langevin4}):
\begin{eqnarray}
\label{langevin9}
-i\omega\zeta_{p}{r}_{px}(\omega) =
-k p^2 {r}_{px}(\omega) &-& 2ip\pi
\lambda_{0}k_{B}T{r}_{py}(\omega)\nonumber\\
 & & ~~~+ f_{px}(\omega)\\
\label{langevin10}
-i\omega \zeta_{p}{r}_{py}(\omega) =
-k p^2{r}_{py}(\omega) &+& 2ip\pi
\lambda_{0}k_{B}T{r}_{px}(\omega)\nonumber\\
& & ~~~ + f_{py}(\omega).
\end{eqnarray} 
The resulting coupled equation can be easily be solved to calculate
the following correlation:
\begin{eqnarray}
\left< {\bf r}_{p}(\omega)\cdot {\bf r}_{-p}(-\omega)\right> &=& 2
k_{B}T\zeta_{p}\nonumber\\
& &\left[\frac{1}{(\zeta_{p}^2\omega^2 + (k p^2 - 2p\pi k_{B}T\lambda_{0})^2)}\right.\nonumber\\
&+& \left. \frac{1}{(\zeta_{p}^2\omega^2 + (k p^2 + 2p\pi k_{B}T \lambda_{0})^2)}\right],\nonumber\\
\end{eqnarray} 
where the angular brackets represent an average with respect to the
random noise. To calculate the equal time correlation we use
$\left<r_{pi}r_{-pj}\right> = \int (d\omega/2\pi) \left<
  r_{pi}(\omega)r_{-pj}(-\omega)\right>$; the resulting expression is
given by
\begin{equation}
\left< {\bf r}_{p}(t) \cdot {\bf r}_{-p}(t)\right> =
\frac{Nb^2/2}{(p^2\pi^2 - \lambda_{0}^2N^2b^4/16)}.
\end{equation}  
The mean square distance of a chord starting from ${\bf r}(0)$ and
ending at any point $s$ along the chain contour is defined by ${\bf R}^2(s) =
({\bf r}(s)- {\bf r}(0))^2$. In terms of the normal modes it is given
by ${\bf R}^2(s) = 4\sum_{p=1}^{\infty}({\bf r}_{p} \cdot {\bf r}_{-p})\sin^2(p\pi s/N)$. In the limit of $N \rightarrow \infty$ and $p\pi/N
\rightarrow q$, the summation can be converted to an integral to give
\begin{equation}
\label{ete}
{\bf R}^2(s) = 2b^2/\pi \int_{0}^{\infty}dq \sin^2(qs)/(q^2 + g^2b^4/16),
\end{equation}
which is the same as Eq. (5.3) in the paper by Brereton and Butler
\cite{brereton}. In writing the above expression we have taken
$\lambda_{0} = ig$ such that the integral can be done by contour
integration; it has double poles in the complex $g$ plane given by $g
= \pm 4iq/b^2$ \cite{brereton}. The latter implies that when $q =
p\pi/N$, the Lagrange multiplier is given by $\lambda_{0} =
\pm 4p\pi/Nb^2$.  

The integral in Eq. (\ref{ete}) can easily be evaluated and Fourier
transformed with respect to $g$; the result can be averaged over the
equilibrium probability distribution given by $P(A) = \exp(-4\pi
A/Nb^2)$ to give the following expression:
\begin{equation}
\label{ete1}
\left<{\bf R}^2(s)\right>_{A} = \frac{Nb^2}{2}(1 + \alpha^2)
\ln\left(\frac{\alpha^2 + (1+s/N)^2}{\alpha^2 +1}\right),
\end{equation} 
where $\alpha = 2A/Nb^2$. It is interesting to note that in equilibrium 
the probability distribution, which yields $\left< A\right> =
Nb^2/4\pi$, imposes the dominance of the first  mode over all other
modes. In essence it means that the presence of the area constraint
introduces significant deviations from the unperturbed random walk for $A/Nb^2 \gg 1$; in the latter limit the mean square distance is given by $\left<{\bf R}^2(s)\right>_{A} \approx sb^2 ( 1 + s/N)^2/2$. 

\subsection{Dynamics}

As discussed in Section III, the dynamics of the Lagrange multiplier function
$\lambda(t)$, when calculated in a self-consistent way, suggests its inverse
dependence on the first mode of the area, i.e., $-4\pi/Nb^2$. The
latter expression when substituted into Eqs. (\ref{langevin9}) and
(\ref{langevin10}) produces coupled equations in $r_{p}(x)$ and
$r_{p}(y)$, which can easily be solved to give
\begin{eqnarray}
\left< {\bf r}_{p}(\omega)\cdot {\bf r}_{-p}(-\omega)\right> &=& 2 k_{B}T\zeta_{p} \nonumber\\
& &\left[\frac{1}{(\zeta_{p}^2\omega^2 + (k^2p^2(p+1)^2)} \right. \nonumber\\
 &+&\left. \frac{1}{(\zeta_{p}^2\omega^2 + k^2p^2(p-1)^2)}\right].
\end{eqnarray} 
The time-correlation function can be calculated using $\left<{\bf
    r}_{p}(t)\cdot{\bf r}_{-p}(0)\right> = \int \frac{d\omega}{2\pi}
\left<{\bf r}_{p}(\omega)\cdot{\bf r}_{-p}(-\omega)\right> {\rm e}^{i \omega
  t}$. For $p=1$ mode the correlation function is given by
\begin{eqnarray}
\left< {\bf r}_{1}(t)\cdot {\bf r}_{-1}(t^\prime)\right> &=& 
\frac{Nb^2}{16\pi^2}\exp\left(-\frac{8\pi^2k_{B}T |t - t^\prime|}{N^2b^2\zeta}\right)\nonumber\\
&+& \frac{k_{B}T|t - t^\prime|}{N\zeta}. 
\end{eqnarray}
For $ p\neq 1$ modes, on the other hand, the time-correlation function 
is given by
\begin{eqnarray}
\left< {\bf r}_{p}(t)\cdot {\bf r}_{-p}(t^\prime)\right> &=&
\frac{Nb^2}{4\pi^2(p^2-1)}\exp\left(-\frac{4\pi^2p^2 k_{B}T|t -
    t^\prime|}{N^2b^2\zeta}\right)\nonumber\\
& &\left[\cosh\left(\frac{4\pi^2pk_{B}T|t - t^\prime|}{N^2b^2\zeta}\right)\right.\nonumber\\ 
& & \left. ~~~+ \frac{1}{p}\sinh\left(\frac{4\pi^2pk_{B}T|t - t^\prime|}{N^2b^2\zeta}\right) \right].
\end{eqnarray}
The radius of the ring at any point $s$ along the curve is given by ${\bf
  R}(s,t) = {\bf r}(s,t) - {\bf r}_{0}(t)$, where ${\bf r}_{0}(t)$ is the
position of the centre of mass at time $t$ defined as ${\bf r}_{0}(t) =
\frac{1}{N}\int_{0}^{N} ds {\bf r}(s,t)$. The time correlation of the radii at
two different times in terms of the normal modes is given by $\left<{\bf
    R}(s,t)\cdot{\bf R}(s,t^\prime)\right> = 4\sum_{p=1}^{\infty}\left<{\bf
    r}_{p}(t)\cdot{\bf r}_{-p}(t^\prime)\right>$. Since the relaxation time
associated with the first mode, that is $\tau = N^2b^2\zeta/8\pi^2k_{B}T$, is
the longest relaxation time of the correlation function, it corresponds to the
rotational relaxation time \cite{doi}. For all times greater than the
relaxation time, that is $t \gg \tau$, the dynamics is dominated by the first
Rouse mode, which is diffusive. The mean square displacement of the radius
${\bf R}(N,t)$ is given by
\begin{equation}
\left<({\bf R}(N,t) - {\bf R}(N,0))^2\right> = \frac{8k_{B}Tt}{N\zeta}, 
\end{equation}
where at long times it corresponds to the rotational diffusion of the ring;
the rotational diffusion constant is given by $D_{rot} = 2k_{B}T/N\zeta$. The
mean square displacement of the centre of mass, on the other hand, is
estimated by the $p=0$ mode, i.e.  $\left<({\bf r}_{0}(t) - {\bf r}_{0}(0))^2\right> = \frac{4k_{B}Tt}{N\zeta}$; at long times the self
diffusion constant of the centre of mass of the ring is given by $D_{0} =
k_{B}T/N\zeta$ \cite{doi}. The rotational diffusion constant is, therefore,
twice of the diffusion constant of the centre of mass of the ring. It is
not surprising, however, that both the diffusion constants have Rouse type
scaling since the only way the diffusion constant can depend on the
system properties is through $D \propto k_{B}T/\zeta
(b^{2}/A$)\cite{doi}. The area is given by $A \propto (1/b^2N)$; the
diffusion constants for the $p=0$ diffusion and the area preserving $p=1$ rotation must, therefore, have the same Rouse type scaling.

Since the area constraint couples the $x$ and $y$ modes, the cross
correlations, i.e. $\left< r_{px}(t)r_{-py}(t^\prime)\right>$, turn out to be
non-zero. The equal time correlation amounts to estimating the average of
$A_{p}(t)$, which can easily be evaluated using Eq. (\ref{area2}). The
summation over $p$ modes defines the average area, which is given by
$\left<A(t)\right> = Nb^2/4\pi$.

The dominance of the diffusive first mode has been seen in a very different
context of grafted polymer brushes \cite{joanny}. In the latter study the
distribution of the chain ends is accounted for by self-consistent-field theory
which determines the mean-field potential seen by each monomer
self-consistently \cite{milner}. In dynamics it amounts to the dominance of
the first mode, which does not relax, and is associated with the diffusion of
the chain end in brushes. In the present study although the imposition of area
constraint through the time dependent Lagrange multiplier and its
determination from the dynamical equations is of very different nature, the
dominance of the first (Rouse) mode, as also seen in the dynamics of
polymer brushes, is interesting.

\section{Conclusions}

In summary, the present theory illustrates the effects of a relatively
simple global constraint on the stochastic dynamics of non interacting
polymer chains. We show that the presence of the area
constraint couples the dynamics of the Cartesian components of the spatial
vector ${\bf r}(s,t) = (r_{x}(s,t), r_{y}(s,t))$ through the time dependent
Lagrange multiplier. The time dependence of Lagrange multiplier can be
determined as a dynamical response to the initial perturbation of area modes
such that the condition $dA(t)/dt =0$ is satisfied at all times. At very short
times, that is $t \ll \tau$, the Lagrange multiplier decays slowly with an
inverse dependence on time, where $\tau = N^2b^2\zeta/8\pi^2k_{B}T$. For long
times, that is $t \gg\tau$, the Lagrange multiplier is mainly determined by
the inverse of the first mode of the area, which on an average is a random
walk about a circle.  As opposed to the relaxation of the internal modes of a
ring without constraint, the area constraint changes the nature of relaxation
of these modes dramatically. At long times the dynamics is completely
determined by the first (Rouse) mode which does not relax. The first mode is associated with the rotation of the ring; the
mean-square displacement of the radius at long times is found to be diffusive.
 
Although the idea was to present a general theory for the
dynamics of area preserving systems, the present
approach can serve as a simple description to understand some of the dynamical aspects of polymer systems
under topological constraints, entanglements and two-dimensional
vesicles --- systems where the area conservation has to be explicitly 
accounted for.  

\acknowledgments

The authors appreciate earlier useful discussions with M.G. Brereton
about this problem. Discussions with Prof. A. Grosberg,
M. Otto and R. Adhikari are gratefully acknowledged.

\end{document}